**Submitted to Journal of Instrumentation**

# Micromegas operation in high pressure xenon: charge and scintillation readout

C. Balan[1], E.D.C. Freitas[1], T. papaevangelou[2], I. Giomataris[2], H. Natal da Luz[1], C.M.B. Monteiro[1], J.M.F. dos Santos[1*]

[1]Instrumentation Centre, Physics Department, University of Coimbra, 3004-516 COIMBRA, Portugal

[2] IRFU, Centre d'Études Nucléaires deSaclay (CEA-Saclay), Gif-sur-Yvette, France

E-mail: jmf@gian.fis.uc.pt

Abstract

The operational characteristics of a Micromegas operating in pure xenon at the pressure range of 1 to 10 bar are investigated. The maximum charge gain achieved in each pressure is approximately constant, around 4 x $10^2$, for xenon pressures up to 5 bar and decreasing slowly above this pressure down to values somewhat above $10^2$ at 10 bar. The MM presents the highest gains for xenon pressures above 4 bar, when compared to other micropattern gaseous multipliers. The lowest energy resolution obtained for X-rays of 22.1 keV exhibits a steady increase with pressure, from 12% at 1bar to about 32% at 10 bar. The effective scintillation yield, defined as the number of photons exiting through the MM mesh holes per primary electron produced in the conversion region was calculated. This yield is about $2\times10^2$ photons per primary electron at 1 bar, increasing to about $6\times10^2$ at 5 bar and, then, decreasing again to $2\times10^2$ at 10 bar. The readout of this scintillation by a suitable photosensor will result in higher gains but with increased statistical fluctuations.

Keywords: Micropatterned Gas detectors, Micromegas, Xenon, High-pressure, Neutrino, Double Beta decay,

* Corresponding author

## 1 Introduction

The "*NEXT - Neutrino Experiment with a Xenon TPC"* collaboration proposes a new concept for neutrino-less double-beta decay search, based on a Time Projection Chamber (TPC) filled with high pressure gaseous Xenon (100 kg at 10 bar and room temperature operation) [1]. The TPC could be based on Electroluminescence for signal amplification, with excellent energy resolution and tracking capabilities.

The noble gas (Xe) has two important functions, as source and detector, both acting simultaneously. Xenon is the only one, among the noble gases, that has a $\beta\beta$ decaying isotope, $^{136}Xe$, and whose natural abundance is rather high, 9%, which can be enriched by centrifugation at a reasonable cost. There are no other long-lived radioactive isotopes and the $Q_{\beta\beta}$ value, 2480 keV, is acceptably high. In addition, xenon can be easily re-circulated and continuously purified through getters.

The TPC energy resolution is essential not only to reduce the tail of the $2\nu\beta\beta$ spectrum from overlapping the region of interest of the $0\nu\beta\beta$ spectrum, but also to prevent the contamination of the region of interest by the most severe background from 2.6 MeV gamma rays from $^{206}Tl$ and 2.4 MeV gamma rays from $^{214}Bi$. On the other hand, the tracking capability aims to further reduce external backgrounds by identifying the unique topological signature of the $0\nu\beta\beta$ events, double electron track, in opposition to a single electron track resulting from gamma interactions. The optimization of these two features is crucial for such experiment with very low event rates and high background levels.

The proposed detector design, called SOFT approach, is based on a specific readout for both tracking and energy measurement [2]. Primary ionization signals are amplified by means of electroluminescence amplification in a confined region of the TPC, the scintillation region. Electroluminescence photons emitted towards the hemisphere of the anode can be used for tracking, while photons emitted in the opposite direction can be detected with a series of PMTs mounted behind the cathode, for energy measurement and $t_0$ determination (by means of primary scintillation readout).

While for the energy and $t_0$ readout plane PMTs are the best choice, for the tracking plane different options can be considered. MPPCs (multi-pixel photo-counters) [3], APDs (avalanche photodiodes) [4] and Si-PMs [5] are options under study for the electroluminescence readout. In addition, the use of Micromegas [6] for the tracking plane is also under consideration [7]. In this case, the primary electrons will be guided to the Micromegas (MM) after crossing the scintillation region, undergoing charge avalanche in the MM gap for signal amplification. On the other hand, if the energy resolution obtained by the Micromegas operating in xenon can be as good as that obtained in other gas mixtures [7,8], i.e. close to the intrinsic energy resolution, then the MM can also present an alternative to the energy readout plane, merging in a single plane the tracking and energy readout and resulting in a significant reduction in detector complexity and cost.

However, the performance of micropatterned electron multipliers operating at high pressures is limited. For GEMs, THGEMs and MHSPs operating in heavy noble gases, Xe and Kr, the maximum achievable gain decreases with pressure, together with an increase of the statistical fluctuations [9-12]. This is due to the fact that the maximum applied voltage does not increase as fast as the pressure and, consequently, the maximum achievable reduced electric field decreases with pressure. Such studies in xenon have not yet been carried out for MM. In addition, the scintillation produced in the electron avalanches in the MM gap will be superimposed with that produced in the scintillation region, due to the large dimensions of the primary electron cloud. This may jeopardize the detector energy resolution if its relative amount is significant.

In this work, we study the performance of Micromegas operating in pure xenon at high pressures. The charge gain, the scintillation yield (i.e. the number of photons leaving the MM per primary electron produced in the conversion gap) and the associate statistical fluctuations are studied as a function of the MM biasing voltage for different pressures, in the range of 1 to 10 bar. The scintillation produced in the MM is readout by means of a large area avalanche photodiode (LAAPD), placed in front of it.

## 2 Experimental set-up

The MM and the LAAPD were accommodated in a stainless steel chamber. The schematic layout of the MM and the LAAPD is depicted in Fig.1. The chamber has a cylindrical shape with 100 mm in diameter and 49.5 mm in height. The MM has an active area of ~ 40 mm in diameter, a gap of 50 μm and its mesh has 25 μm diameter holes. The LAAPD is an API Deep UV model [13], with an active area 16 mm and its encapsulation is perforated with holes in order to have the same pressure in both sides of the Si wafer. The MM backplane was fixed to the chamber scintillation window. A stainless steel mesh (80 μm diameter wires with 900 μm spacing) was placed between the MM and the LAAPD in order to establish an uniform electric field in the conversion/drift region. The stainless steel mesh and the MM mesh were kept at negative voltages while the MM induction plane was kept at zero volts. The LAAPD enclosure and the chamber body were grounded. The conversion/drift region gap was 7.0 mm thick and the region between the LAAPD enclosure and the stainless steel mesh was chosen to be thick, 7.8 mm, in order to keep the reduced electric field in this region below the xenon scintillation threshold (~ 0.8 kV $cm^{-1}bar^{-1}$ [14]).

The Micromegas [6] is a double stage parallel plate avalanche counter with a narrow multiplication gap (25-150 μm, 50-70 kV/cm), located between a thin metal grid (micromesh) and the readout electrode (strips/pads of conductor printed on an insulator board). The distance homogeneity between the anode and the grid mesh is preserved by using spacers from insulating material. The small amplification gap is a key element in Micromegas operation, giving rise to excellent spatial and time resolution: 12 μm spatial (limited by the pitch of micromesh) and 300 ps time resolution are achieved with single electron signal.

Microbulk Micromegas are manufactured with a novel technique, based on kapton thin-foil etching technology [15]. A thin photoresistive film is laminated on top of the kapton foil and it is insolated by UV light to produce the required mask. The copper is then removed by a standard lithographic process, the non-insulated places producing a pattern of a thin mesh. The polyimide is then etched and partially removed in order to create tiny pillars in the shadow part of the mesh below the copper mesh. The result is an "all-in-one" detector with improved characteristics such as uniformity, stability and material radiopurity. Thus, the achieved maximum gain and energy resolution are further improved compared to the traditional Micromegas, while it is possible to construct detectors with multiplication gaps of 25 or even 12.5 μm which are better performing in high gas pressures.

A 2-mm collimated X-ray beam originating from a $^{109}Cd$ source (22.1 and 25.0 keV Ag $K_\alpha$ and $K_\beta$ fluorescence X-rays, respectively), irradiates the conversion/drift region through the chamber radiation window and through the MM. The primary electron clouds induced by the X-rays in the conversion gap were focused under a drift field $E_{drift}$ into the MM holes and multiplied in the gap, with the charge signal being collected in the MM induction electrode. A great number of VUV photons (λ ~ 172nm) are produced along the charge avalanche as a result of the gas de-excitation processes. Part of these photons leaves the MM trough the mesh holes and reaches the LAAPD active area and the corresponding electric signal is amplified in the photodiode. Therefore, we have two independent readout channels: one for the MM induction plane, which we call charge channel, and the other for the LAAPD anode, which we call scintillation channel.

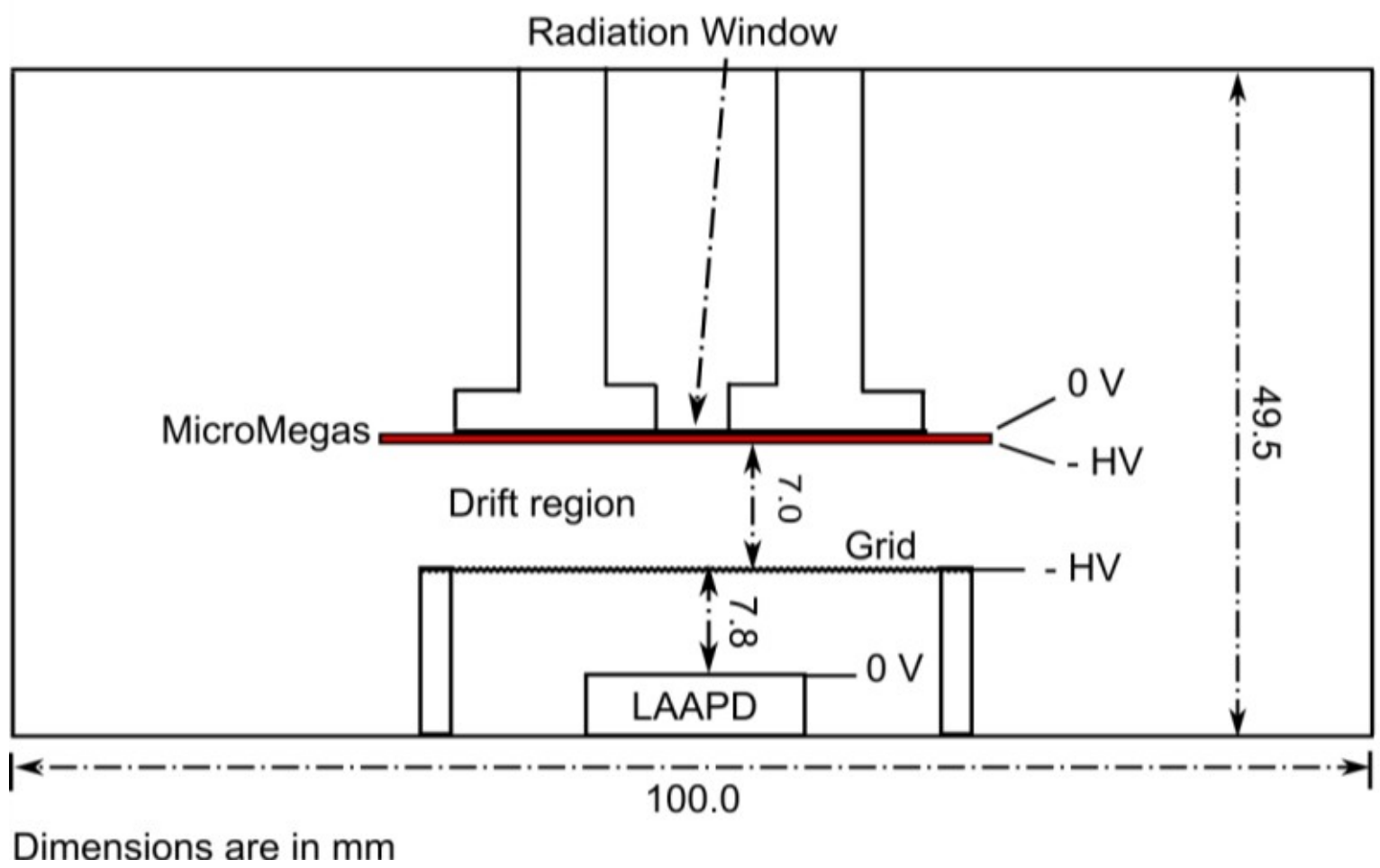

Figure 1- Schematic layout of the MicroMegas detector used in this work.

Through this experimental work, the LAAPD biasing was set to a safe value of 1650 V, corresponding to a charge amplification gain of about 30 [16,17]. High performance is reached, even for such small photosensor gains, as a result of both high scintillation amplification and high conversion-efficiency of xenon scintillation into charge in the LAAPD [18]. The LAAPD was used to detect simultaneously the secondary scintillation produced in the MM and the incident X-rays, which are used as a reference for determining the absolute number of charge carriers produced by the scintillation detected in the LAAPD and, hence, the number of VUV-photons hitting the LAAPD, given its quantum efficiency.

The chamber was pumped down to ~$10^{-5}$ mbar by a turbo-molecular pump and filled with Xenon at pressures from 1 to 10 bar. The pressure was kept constant during each set of measurements. The xenon is continuously purified, circulating by convection through non-evaporable getters (SAES St 707), which are kept to a stable temperature of ~130ºC.

The charge signals of both MM induction plane and LAAPD were collected by charge-sensitive preamplifiers, being further amplified and shaped with linear amplifiers and pulse-height analyzed with multi-channel analyzers. For peak amplitude and energy resolution measurements, pulse-height distributions were fitted to a Gaussian function superimposed on a linear background to determine the centroid and the full width at half maximum (FWHM). Both electronic chains were calibrated by means of a precision pulse generator and a known capacitance coupled to the preamplifier input.

## 3 Experimental results

Typical pulse-height distributions obtained for the $^{109}$Cd X-rays are presented in Fig.2, for filling pressures of 2, 6 and 10 bar and for the charge and the scintillation readout channels. The pulse-height distributions exhibit the peak resulting from the Ag K-fluorescence emitted by the $^{109}$Cd X-ray radioactive source, the peak resulting from the Cu K-fluorescence resulting from the interactions of the X-rays in the MM copper electrodes, and the electronic noise tail in the low energy range. The energy resolution is not enough to separate the Ag $K_{\alpha}$ and $K_{\beta}$ lines. A good separation has been observed [19] in a different set-up using an Argon mixture at atmospheric pressure. We do not know whether this degradation is due to the different gas or pressure or the quality of the read-out element. We intent to repeat these measurements using a selected MM detector.

For the scintillation readout channel an additional peak resulting from the Ag K-fluorescence direct interactions in the LAAPD is present in the pulse-height distributions up to 7 bar. Above that pressure this peak is inside the electronic noise tail.

On the other hand, the position of the peak depends only on the LAAPD biasing voltage and the peak is present even for null voltage difference across the MM gap.

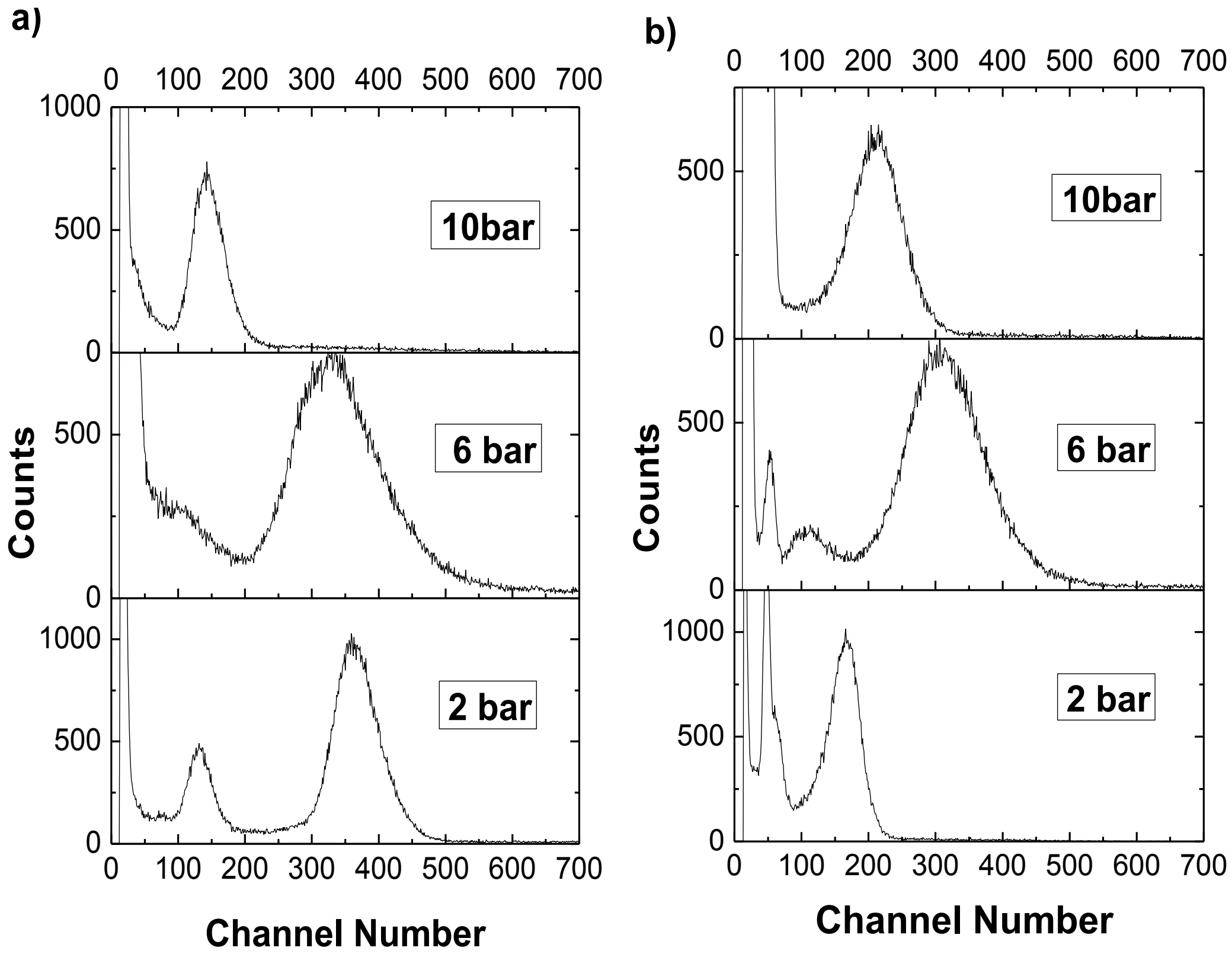

Figure 2- Typical pulse-height distributions obtained for the $^{109}$Cd X-ray source; a) for the charge readout channel and b) for the scintillation readout channel.

We have evaluated the effect of the electric field intensity of the drift region on the charge gain in order to establish good operational conditions. In Fig.3, we depict the MM relative amplitude as a function of reduced electric field in the drift region. We observed that the charge gain and energy resolution are fairly constant over a wide range of electric field values, down to very low values. Even for null or reversed electric fields in the drift region, the MM charge gain is significant. This is due to the penetration of the very intense electric field present in the MM gap into the shallow drift region of this detector. For high values of drift electric field, the MM charge gain may decrease as a consequence of the decrease of the primary electron transmission through the MM mesh, which depends on the ratio between the electric fields in the drift region and in the MM gap. The drift voltage plateau is wider at higher gas pressures. This effect could be attributed to the lower diffusion coefficient which prevent electron losses at the mesh. Note that electron transmission could be optimized by using a more transparent mesh. Such mesh is currently under development.

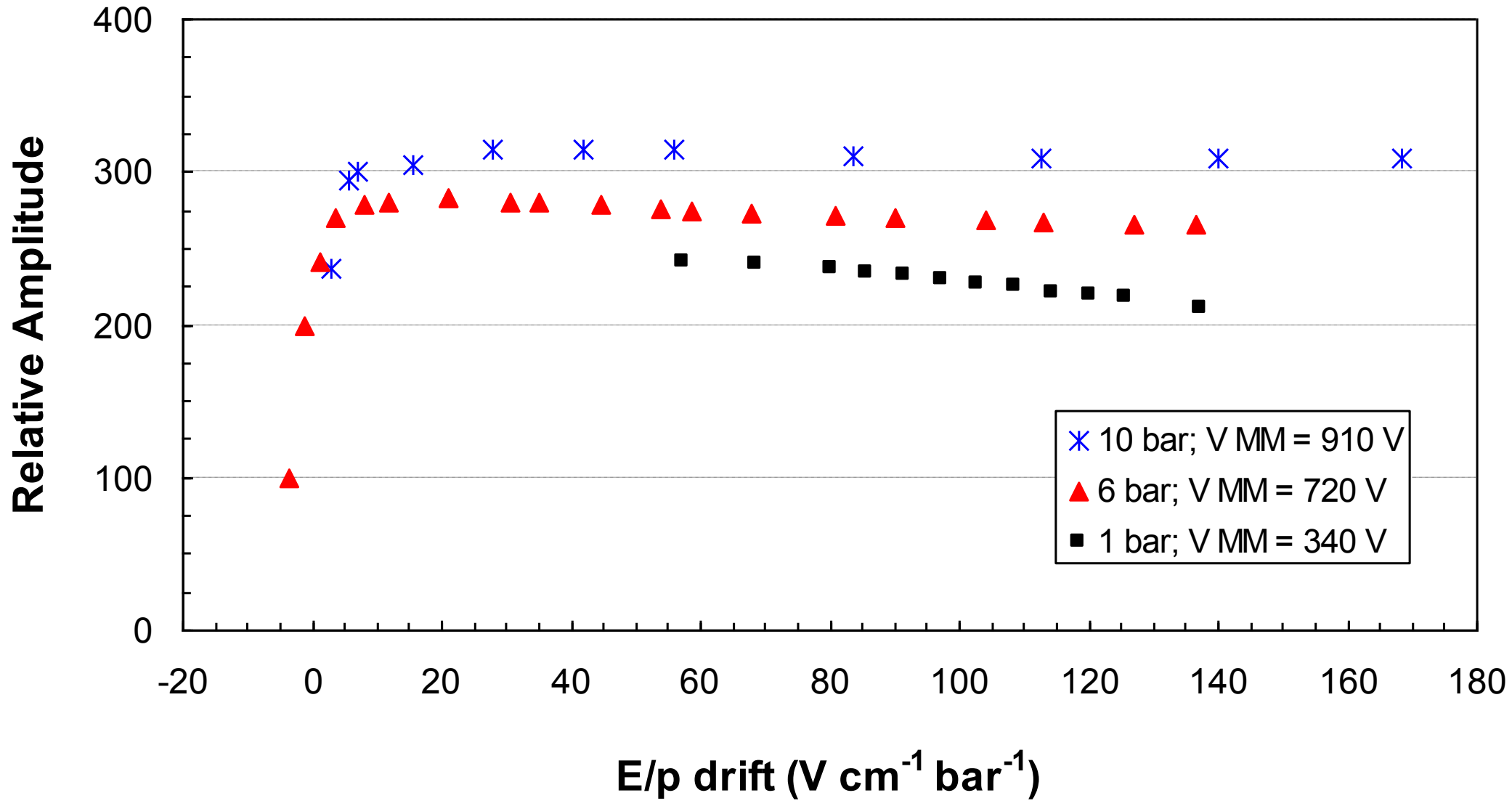

Figure 3- MM relative amplitude as a function of reduced electric field in the drift region.

### 3.1 Gain

Figure 4 depicts typical gain curves obtained with the present MM as a function of MM biasing voltage, for both charge and scintillation readout channels, Figs. 4a) and 4b), respectively, and for different filling pressures. The MM biasing voltage was increased until a first discharge occurred and the run was ended. For the scintillation readout channel, the photosensor charge amplification gain was 30. From this value and from the quantum efficiency of the LAAPD it is possible to determine the number of photons hitting the LAAPD, as it will be discussed ahead.

Figure 4a) shows that the maximum absolute gain of the charge readout channel presents only a small dependence with pressure, increasing within a factor of two up to 5 bar and decreasing slowly above this pressure. The maximum gain achieved at 10 bar is still above 100 and it is only four times less than the highest gain. This behavior is in opposition to other micropattern gas electron multipliers, where the charge gain of electron avalanches presents a fast decrease with pressure, with a gain reduction of a few orders of magnitude.

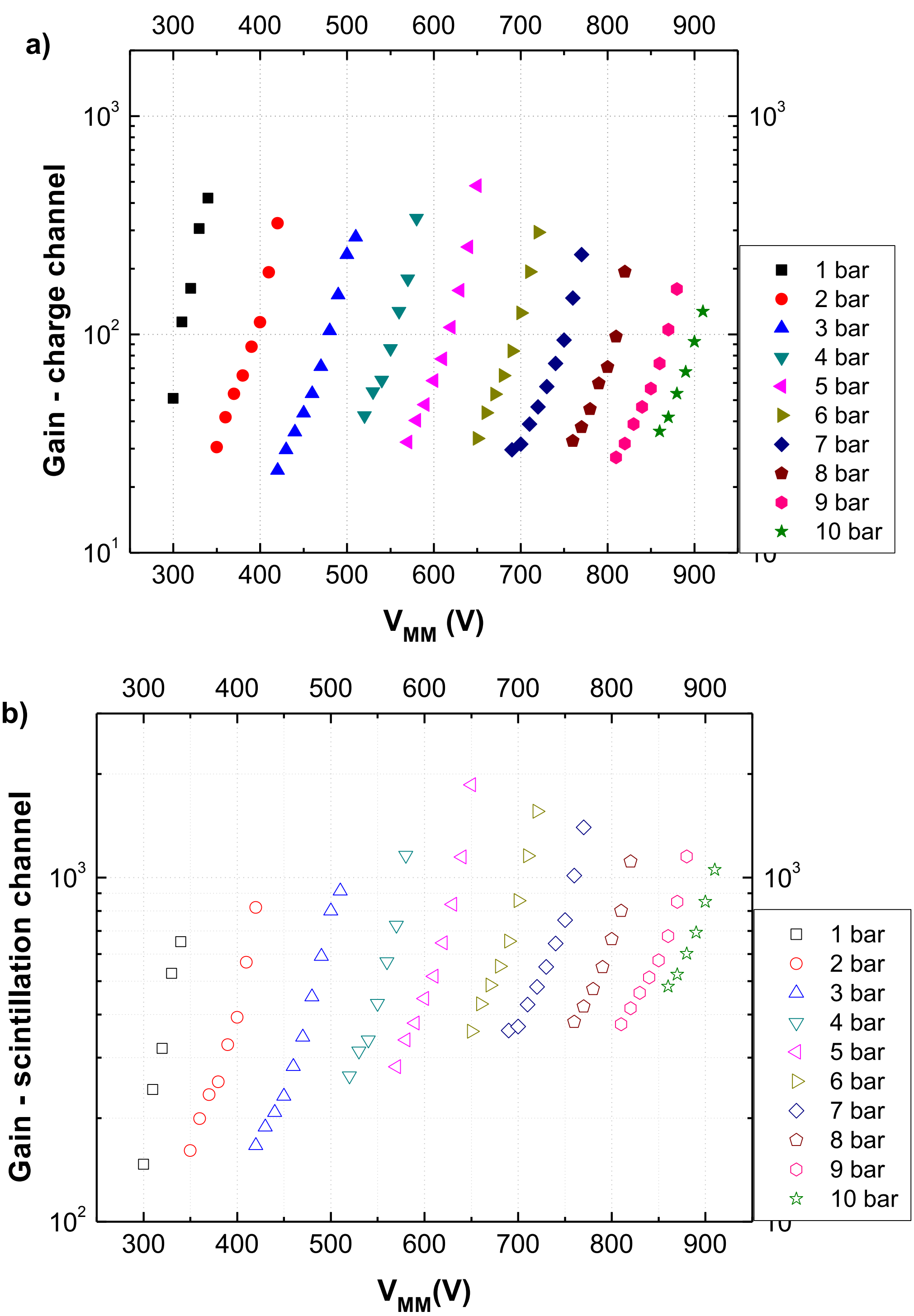

Figure 4- MM charge gain (a) and total charge amplification gain of the scintillation channel readout (b), as a function of MM biasing voltage.

These results evidence a notable characteristic of MM and its potential to be used for high pressure operation. This is clearly shown in Fig.5 where the maximum achieved gain in MM is depicted as a function of pressure, together with the maximum gains obtained for triple-GEM [9], MHSP [10], GEM [11], THGEM [12]. Although at 1 bar the MM is the microstructure presenting the lowest gains, in xenon, as the pressure increases the MM gain becomes higher than that obtained with the other microstructures, Fig.5.

The maximum gain achieved in the scintillation readout channel presents an even smaller dependence with pressure, increasing by a factor of 3 up to 5 bar and, then, decreasing within a factor of two up to 10 bar, as shown in Fig.4b) and Fig.5. The scintillation produced in the electron avalanches from GEMs have been also readout by a LAAPD of the same type, presenting a much faster gain reduction with increasing pressure, a factor of 5 in decrease when the pressure increases from 1 to 2.5 bar [20].

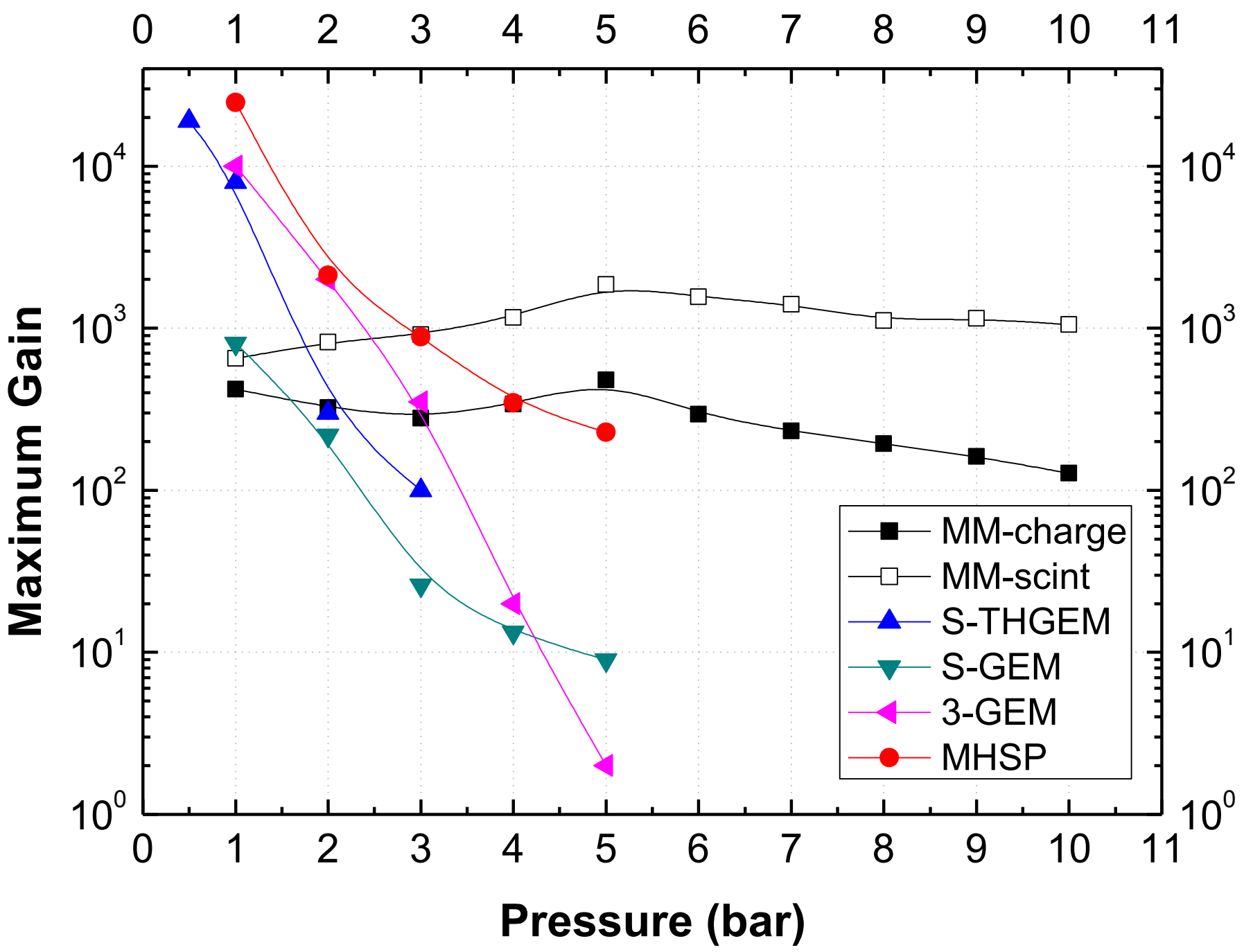

Figure 5 – Maximum gains obtained with the MM for both charge and scintillation readout channels as a function of pressure in the 1 to 10 bar range. For comparison the maximum gains obtained with other micropattern gas electron multipliers are also depicted as a function of pressure: triple-GEM [9], MHSP [10], GEM [11], THGEM [12].

Figure 6 presents the maximum operation voltage that can be applied to the different microstructures as a function of pressure. In single-element multipliers, the maximum applicable voltage steadily increases with pressure, being the MM the microstructure withstanding the lowest voltages, but delivering higher charge gains for pressures above 3-4 bar. Studies on the electron avalanche mechanisms have been performed [21,22]. For Xe, the electron avalanche ionisations and excitations are determined by the electron-impact mechanism, which explains the maximum gain drop for high pressures, as the maximum applied voltages do not increase as fast as pressure, Fig.6.

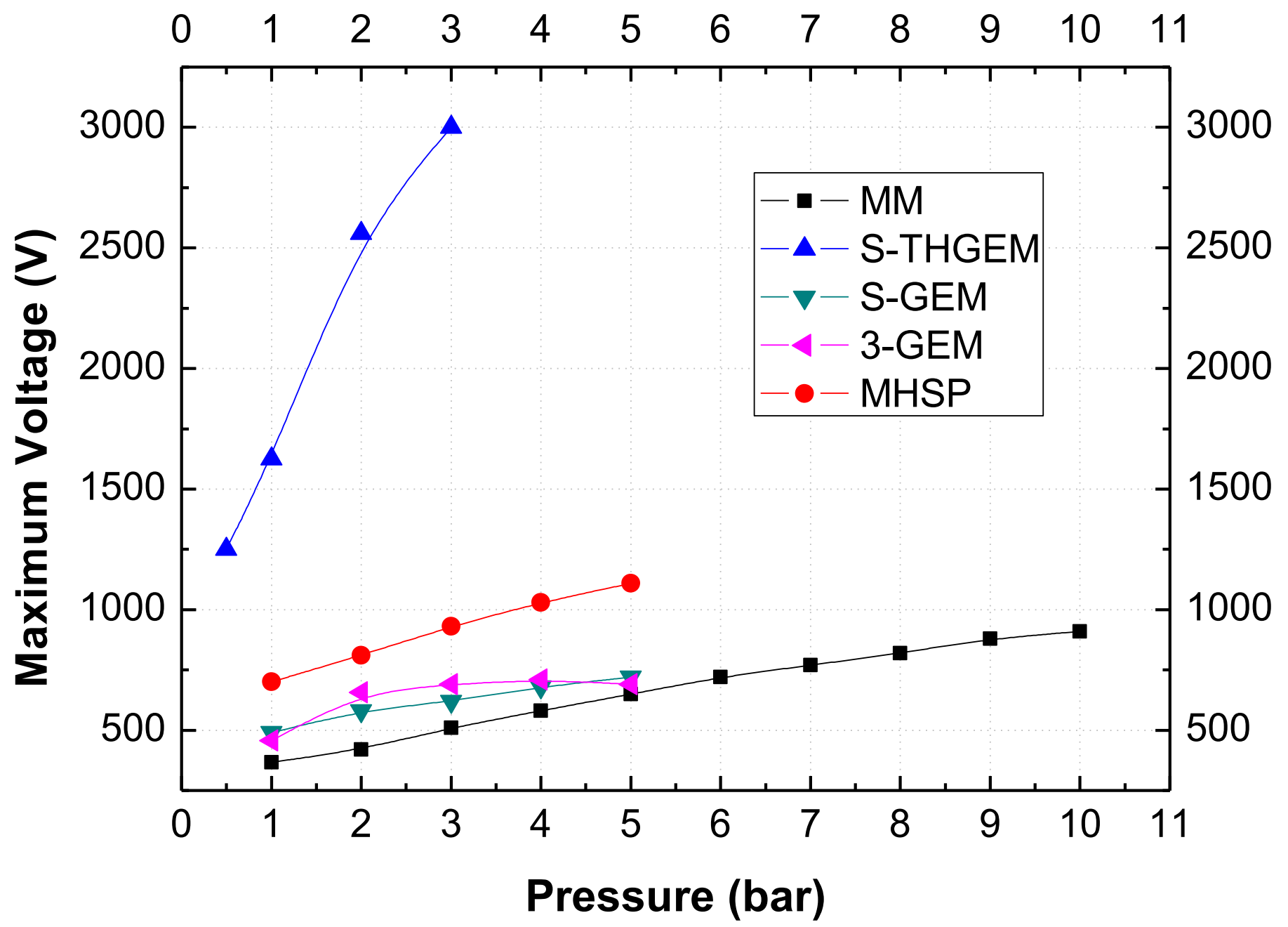

Figure 6 – Maximum operation voltage as a function of the pressure for the different microstructures. For comparison the maximum gains obtained with other micropattern gas electron multipliers are also depicted as a function of pressure: triple-GEM [9], MHSP [10], GEM [11], THGEM [12].

Figure 7 depicts the charge-to-scintillation gain ratio as a function of the MM biasing voltage for the different xenon pressures. The data is consistent with the trend of the reduced electric field in the MM gap. Higher reduced electric fields favor the gas ionization when compared to gas excitation. Therefore, the charge-to-scintillation ratio increases with the MM bias voltage, for each gas pressure, and decreases with pressure, as the maximum applied voltages do not increase as fast as pressure and, consequently, the reduced electric field decreases with pressure. The gain of the scintillation readout channel is, in any case, less than a factor of 10 when compared to the gain of the charge readout channel, in opposition to GEMs and THGEMs, for the same photosensor conditions [21]. This may be due to the small MM mesh hole diameter, which is half of the gap thickness, limiting the amount of scintillation that can exit the MM.

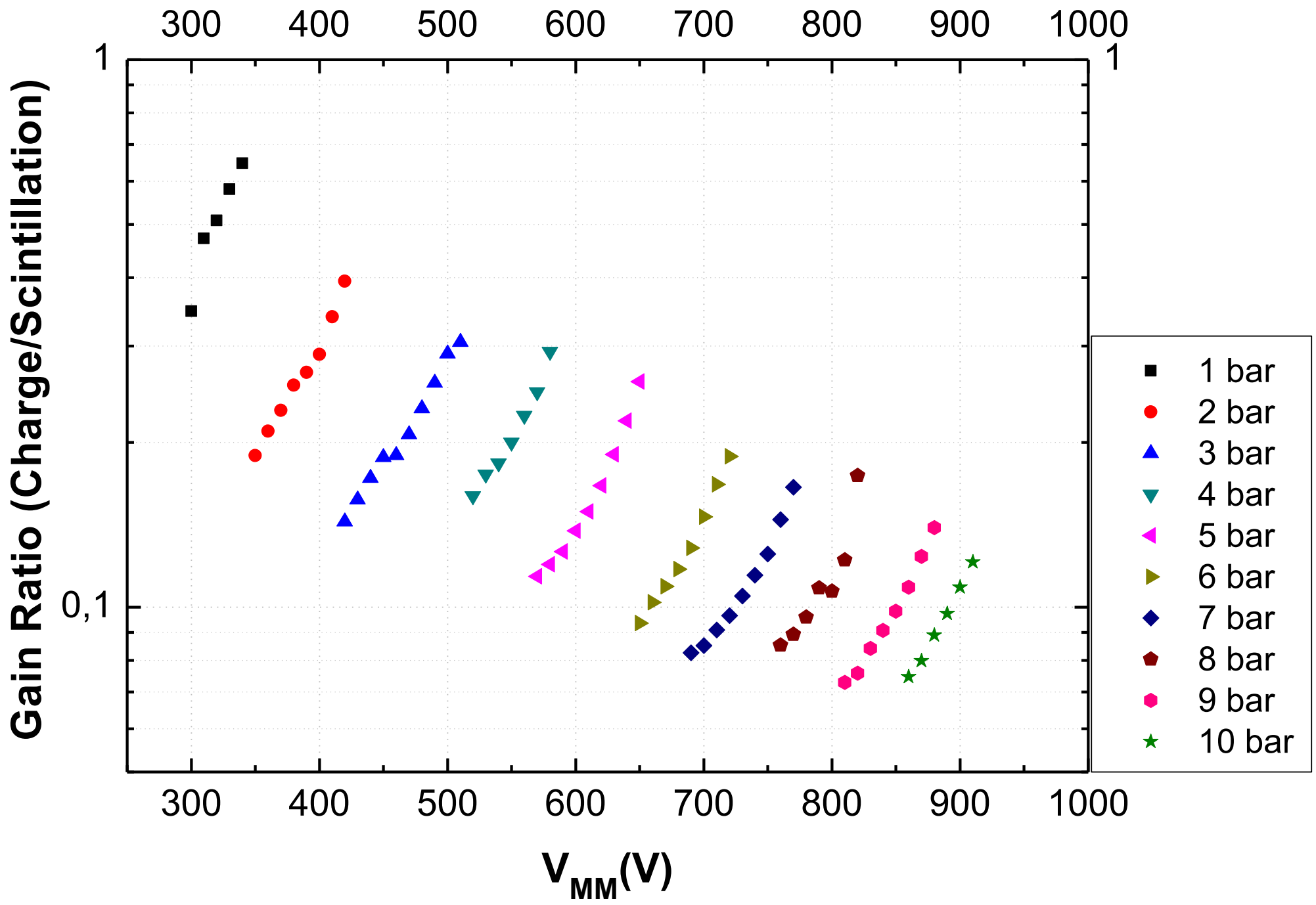

Figure 7 – Charge-to-scintillation gain ratio as a function of the MM biasing voltage for the different xenon pressures.

### 3.2 Energy Resolution

The reduced electric field in the drift region defines the primary electron cloud diffusion and the ratio of drift-to-gap fields defines the electron transfer efficiency through the mesh. Therefore, we have studied the energy resolution as a function of the drift electric field for different xenon pressures. We found that, similarly to the gain, the energy resolution is fairly constant over a wide range of drift reduced electric field values, down to very low values, Fig.8. The penetration of the very intense electric field present in the MM gap into the shallow drift region of this detector is responsible to the efficient focusing of primary electrons into the mesh aperture even at very low drift fields.

Figure 9 depicts the energy resolution obtained for the 22.1 keV X-rays as a function of the MM biasing for the different xenon pressures and for both charge readout and scintillation readout channels, Figs.9 a) and b), respectively. For each pressure, the energy resolution presents a fast decrease as the MM biasing voltage increases, reaching a minimum and then increasing again for the highest voltages. This degradation is due to the onset of ion and/or photon feedback effects in the mesh electrode for the highest avalanche gains. This effect is also seen in the gain curves, which present a supra-exponential increase for the higher voltages. On the other hand, the fast increase in the energy resolution for low biasing voltages is due to the poor signal-to-noise ratio.

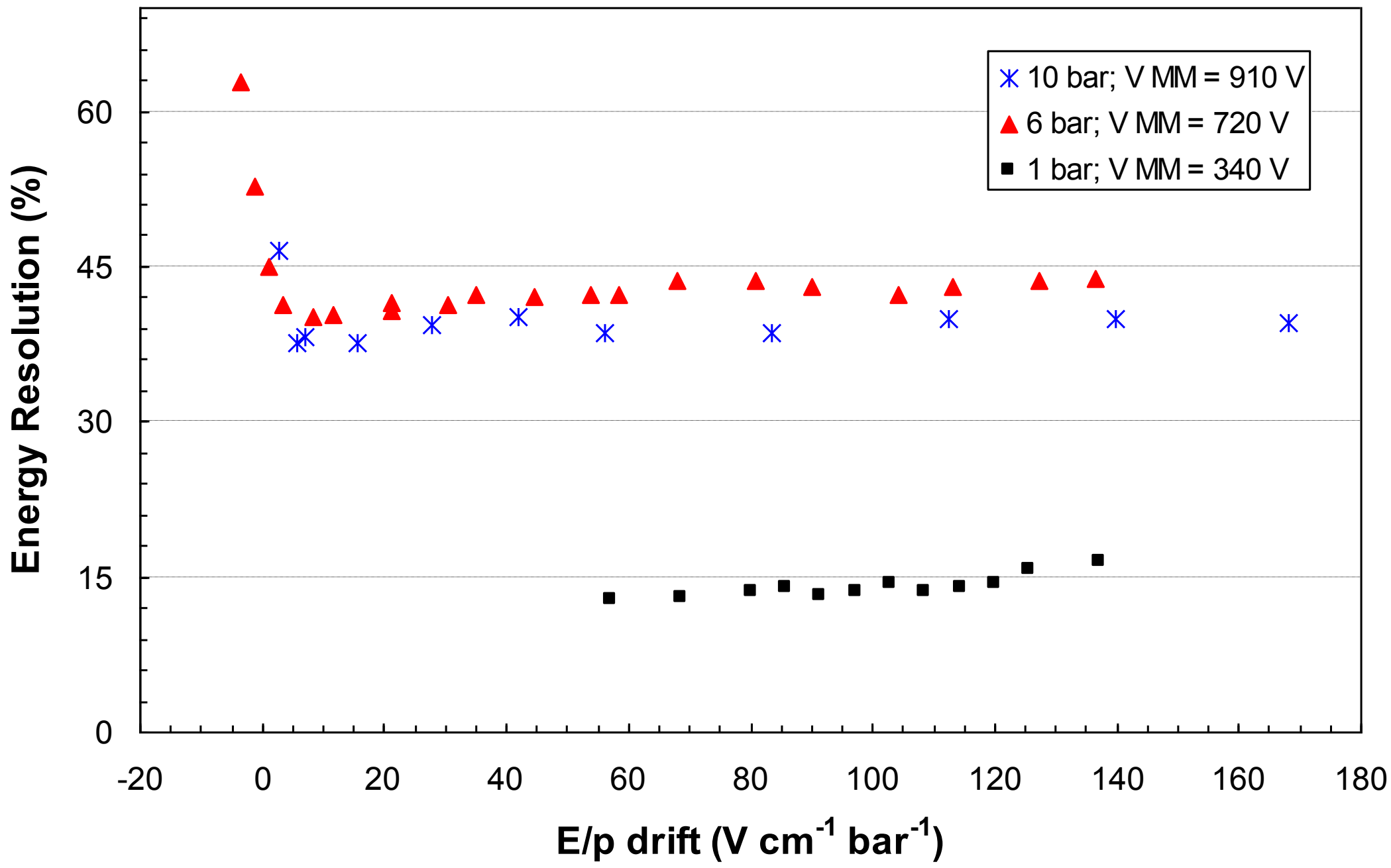

Figure 8 – Energy resolution of the charge readout channel as a function of reduced electric field in the drift region for xenon pressures of 1, 6 and 10 bar.

Fig.9a) shows that the best energy resolution achieved in the charge readout, at each pressure, increases steadily with pressure from 12% at 1 bar to 32% at 10 bar. These energy resolutions are better than those obtained with GEMs and THGEMs. Fig.9b) shows that the energy resolution obtained for the scintillation readout channel is higher than that obtained for the charge readout and presents a different behavior with pressure; the best energy resolution is approximately constant, about 30%, up to 7 bar, degrading to about 40% for the highest pressures.

Fig.10 depicts the signal-to-noise ratio, SNR, as a function of pressure, for the highest gains. As shown, given the low gains achieved in the MM, the SNR is low, below 10, being around 3-4 for pressures above 8 bar. However, a more elaborate system, optimized against noise is needed to further study this issue.

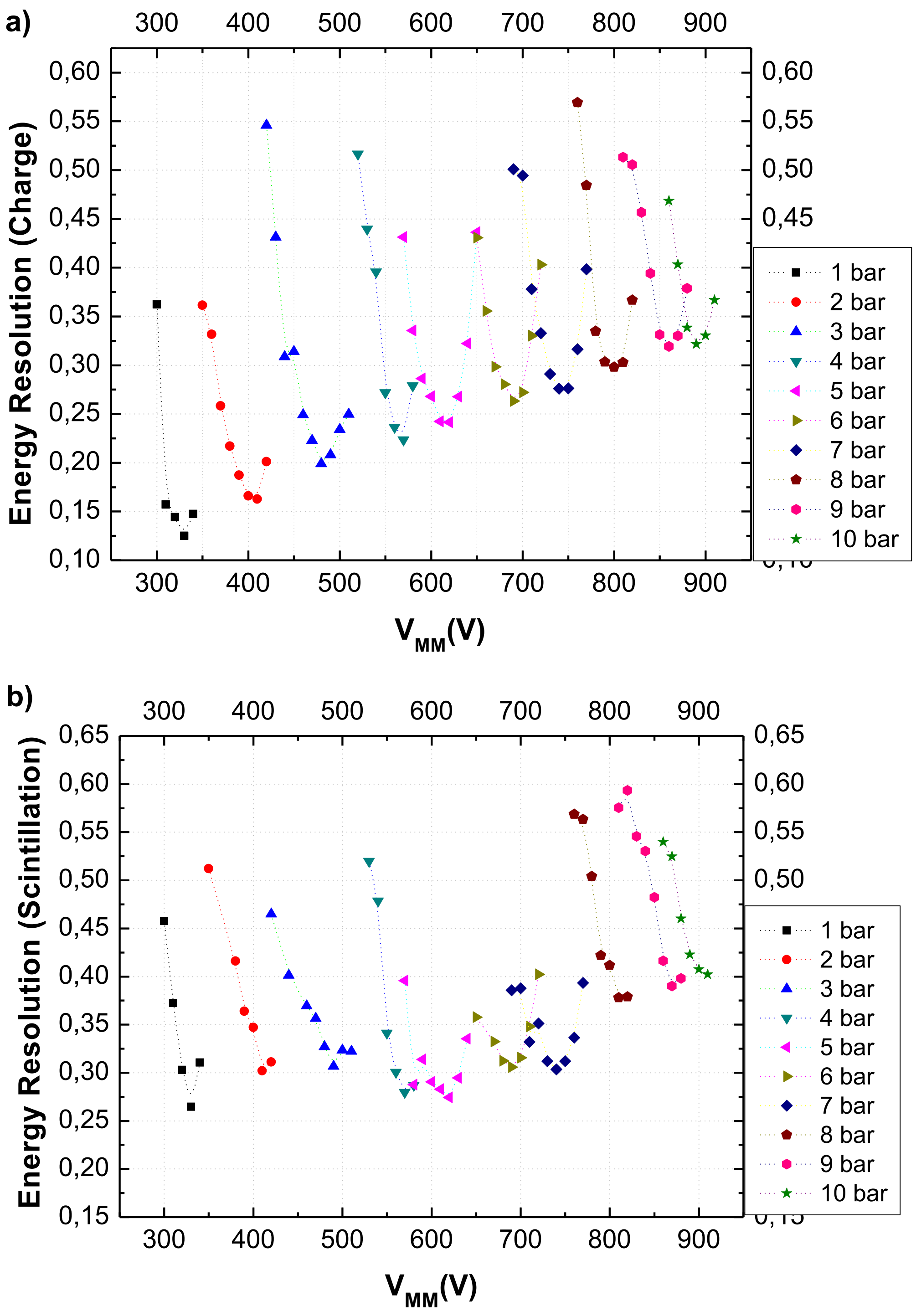

Figure 9 – Detector energy resolution for the 22.1 keV as function of MM biasing voltage and for the different xenon pressures: a) charge readout channel, b) scintillation readout channel.

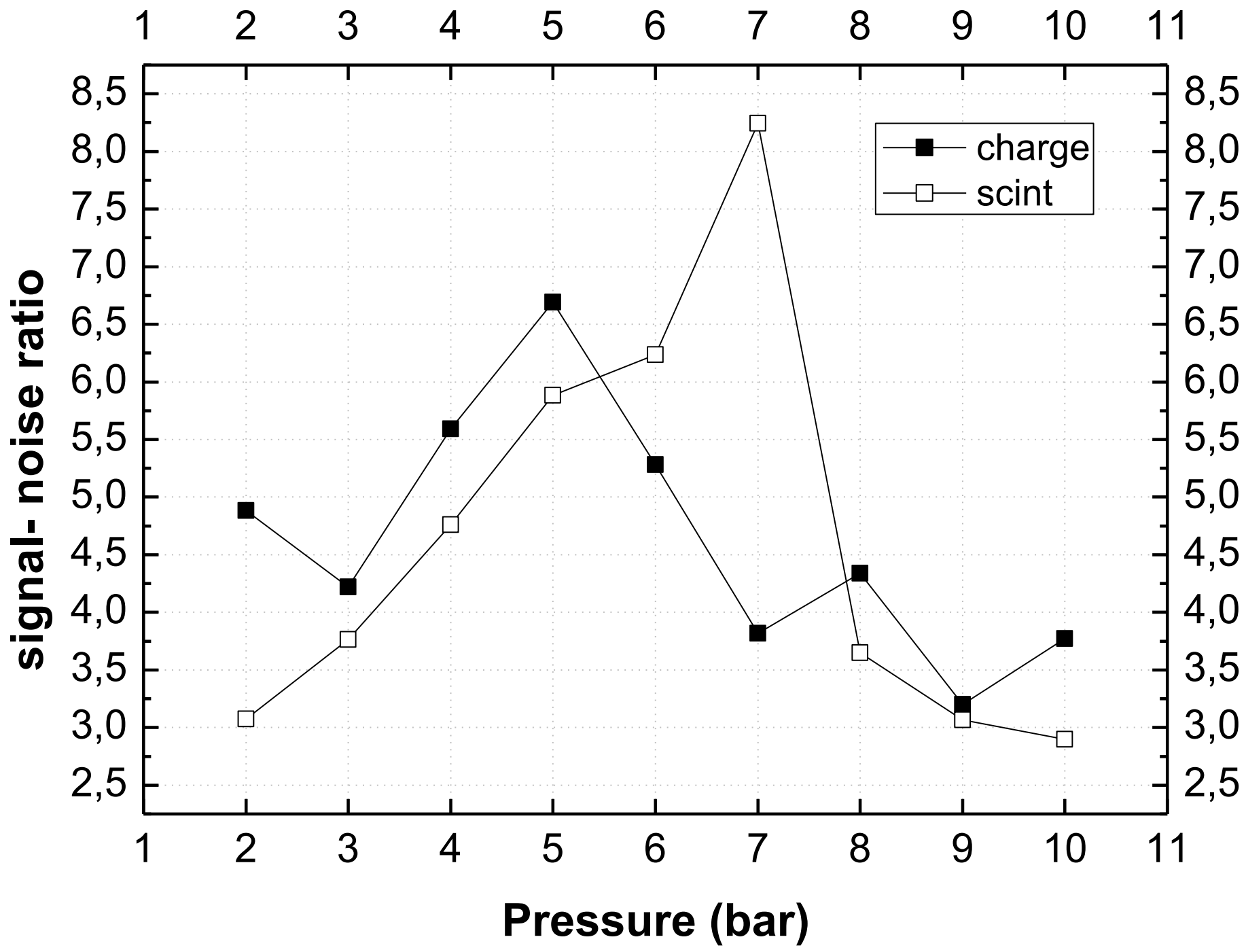

Figure 10 – Signal/Noise Ratio as a function of pressure for maximum applied voltage in the MM gap.

### 3.3 Scintillation yield

We define the effective scintillation yield of MM, $Y_{eff}$, as the number of photons that arise from the MM holes per primary electron produced in the drift region. A certain amount of these photons hit the LAAPD active surface, producing a number of free charge carriers, which are amplified and the resulting charge signal is collected in the LAAPD anode.

The number of photons impinging the LAAPD per X-ray interaction, $N_{uv}$, is related to the effective scintillation yield through

$$Y_{eff} = N_{UV} \times \frac{2\pi}{\Omega_{Sc}} \times \left( \frac{E_x}{w_{E_x}} \right)^{-1} \qquad (1),$$

where $\Omega_{sc}$ is the solid angle subtended by the LAAPD, $E_x$ is the energy of the incident X-ray and $w_{Ex}$ the respective w-value for xenon. In our conditions, the *w*-value for xenon is 21.77 eV for 22.1 keV X-rays [23] and the relative solid angle subtended by the LAAPD is $\Omega_{sc}/2\pi = 0.12$. $N_{uv}$ can be determined through the pulse-height distributions obtained with the scintillation readout channel, by comparing the centroid of the peak resulting from the full absorption of the 22.1 keV X-rays in the drift region, i.e. resulting from the xenon electroluminescence, $A_{Sc}$, with the centroid of the peak resulting from the direct absorption of the X-rays in the LAAPD, $A_X$,

$$N_{UV} = \frac{A_{Sc}}{A_X} \times \frac{N_{e,XR}}{QE} \qquad (2),$$

where $QE$ is the quantum efficiency of the LAAPD, defined as the number of charge carriers produced per incident VUV photon, being 1.1 for 172-nm photons [24,25], and, $N_{e,XR}$ is the number of electron-hole pairs produced by direct absorption of the X-ray in the LAAPD. The latter is determined from the energy of the X-ray and the $w$-value in silicon ($w = 3.62$ eV [26]) and is approximately $6.1\times10^3$ electron-hole pairs for $E_x = 22.1$ keV. The dominating source of uncertainty in the calculated yield is QE, which is estimated with an error of ± 10% [17]. This method has been used to determine the xenon and argon electroluminescence yield for uniform electric fields, below and just above the ionization threshold and the electroluminescence yield produced in the GEM avalanches [27, 28, 20], presenting results that agree with those available in the literature.

The MM effective scintillation yield, is depicted in Fig.11 as a function of MM biasing voltage for the different xenon pressures. The total number of photons released by the MM operating in xenon is about 200 photons per primary electron produced in the drift region, at low pressures in the 1- 3 bar range. This is more than one order of magnitude lower than the total number of photons produced in the GEM avalanches [20]. We believe that this difference is due to the fact the present MM has a mesh with holes having small diameter, 25 μm, being most of the scintillation produced in the last part of the electron path in the gap, i.e. almost 50 micron away from the holes. Therefore, the small average solid angle subtended by the mesh holes reduces significantly the amount of scintillation produced in the MM gap that passes through the holes. For xenon pressures of 10 bar the MM effective scintillation yield is just above 200 photons per primary electron.

The calibration of the electronic chain of the scintillation readout channel allows an independent determination of the MM effective scintillation yield. This calibration allows the calculation of the number of electrons collected in the LAAPD anode per primary electron produced in the scintillation region, $G_{tot}$, i.e. the data presented in Fig.4b). Assuming the LAAPD gain for the 1650 V bias to be about $G_{APD}$ = 30 [16,17], the number of charge carriers produced by the scintillation pulse can be determined by the ratio of these two gains. Therefore, the number of photons impinging the LAAPD per primary electron, $N_{uv,e}$, can be given by

$$N_{UV,e} = QE^{-1} \times \frac{G_{tot}}{G_{APD}} \qquad (3),$$

and the MM effective scintillation yield is obtained from

$$Y_{eff} = N_{UV} \times \frac{2\pi}{\Omega_{Sc}} \qquad (4).$$

The values obtained from Eq.4 are also depicted in fig.11 and are similar to those obtained with the former method. However, the uncertainty in the yield obtained by this method is higher, because of the uncertainty in $G_{APD}$ and in $G_{tot}$, which are larger, than that of $QE$.

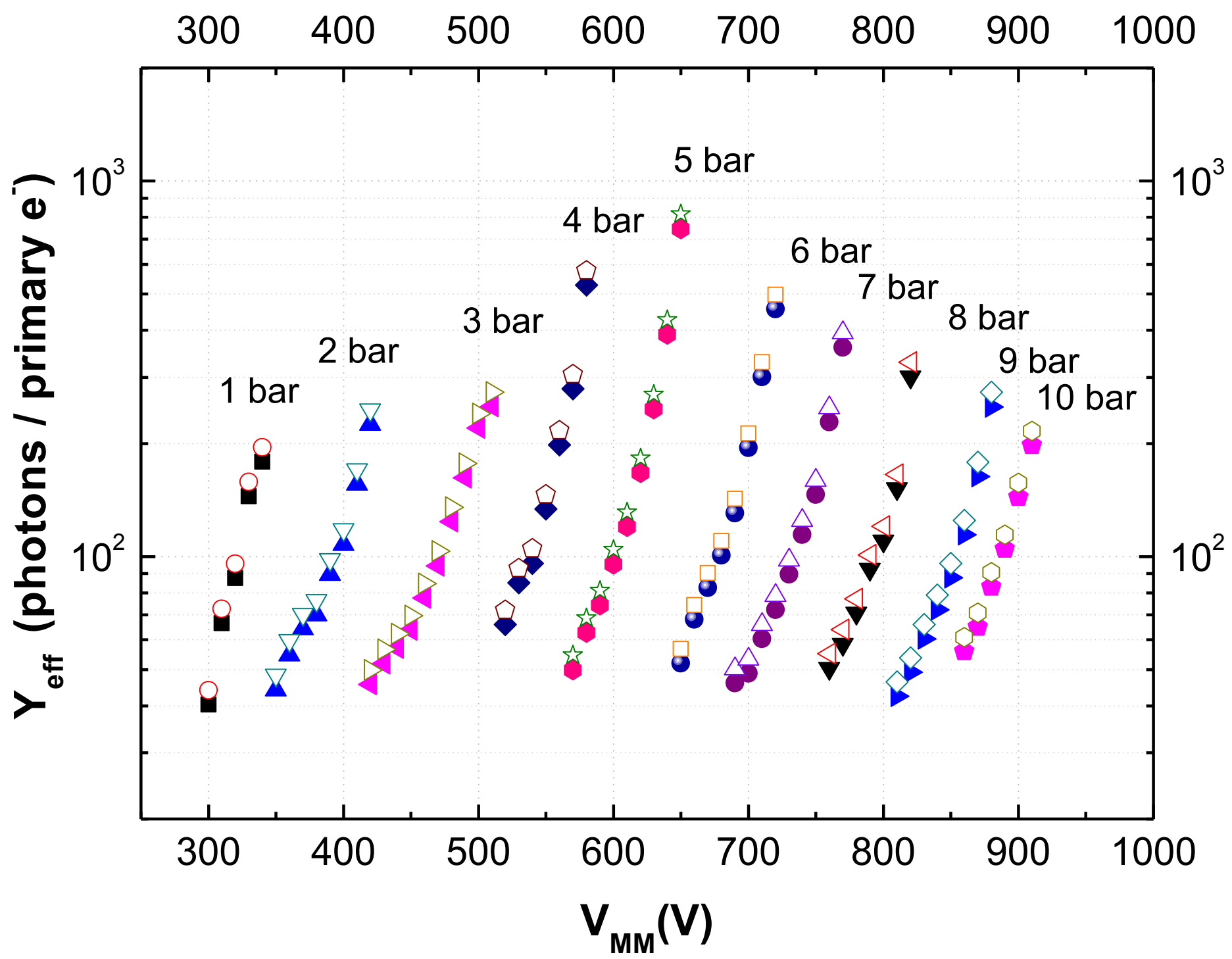

Figure 11 – Effective scintillation Yield, i.e. number of photons emitted from the MM per primary electron produced in the drift region, as a function of MM biasing voltage. Solid symbols: using direct X-ray interactions in the LAAPD as a reference; open symbols: using the gain calibration of the electronic chain and an LAAPD gain of about 30.

## 4 Summary and discussion

We investigated the characteristics of the Micromegas electron multiplier operated in Xe, at pressures ranging from 1 to 10 bar. The results exhibit the remarkable behaviour of MM for high pressure operation. The charge gains are around 4 x $10^2$ for xenon pressures up to 5 bar, decreasing slowly above this pressure down to values above $10^2$ at 10 bar. Although the gains achieved with the MM at low pressures are much lower than those obtained with THGEMs, MHSPs and GEMs, the MM presents the highest gains for xenon pressures above 4 bar. On the other hand, the energy resolution obtained for X-rays of 22.1 keV exhibits a steady increase from 12% at 1bar to about 32% at 10 bar, presenting at 1 bar similar values to those of GEM and substantially better than those of THGEMs. For high pressures the energy resolutions obtained with the MM are better than those obtained with the GEM.

The scintillation produced in the electron avalanches in the MM gap was assessed and the effective scintillation yield, defined as the number of photons exiting through the MM mesh holes per primary electron produced in the conversion region was calculated. This yield is about 2x$10^2$ photons per primary electron at 1 bar, increasing to about 6x$10^2$ at 5 bar and, then, decreasing again to 2x$10^2$ at 10 bar. At 1 bar, the amount of scintillation emitted by the MM is more than one order of magnitude lower than that emitted from GEMs, THGEMs and MHSPs. The readout of this scintillation by a suitable photosensor will result in higher gains but with increased statistical fluctuations.

Concerning the potential application of MM to the NEXT experiment, it is clear that the MM is the only micropatterned structure that can potentially be operated in 10 bar xenon and, therefore, that can be used for the tracking plane of the NEXT TPC. The obtained signal-to-noise ratio shows that the minimum detectable energy can be easily set at few keV, enough for the tracking performance. However, the effective scintillation yield exiting through the holes of the present MM is of the same order of magnitude as the yield produced in the scintillation gap and these two components will be superimposed, due to the large diffusion in the TPC of the primary electron cloud. This will be a drawback in the use of this MM for the tracking plane for it will degrade the energy resolution of the TPC, which should be optimized to the best possibly achievable. The results also show the energy resolution obtained for the charge readout in the MM is worse than that obtained for the readout of the scintillation produced in an uniform field gap, without the presence of charge multiplication, either using a PMT or a LAAPD, which may present resolutions below 6% at 10 bar, for 22.1 keV X-rays [29]. This observation, in favour of scintillation read-out, should be taken into account for the design of the read-out plane of NEXT TPC. However this result must be extrapolated at higher energy in the MeV region using a realistic large size TPC prototype with an appropriate light collection system.

Nevertheless, other MM types aiming to reach much higher gains and much lower energy resolutions in high pressure xenon operation are under implementation [30,31] and will be subject to similar studies to evaluate its potential applicability to the NEXT TPC.

**Acknowledgements**

Financial support is acknowledged from FEDER and FCT, Lisbon, through projects CERN/FP/109324/2009 and PTDC/FIS/103860/2008. HN Luz and EDC Freitas acknowledge grant Nº SFRH/BPD/66737/2009 and SFRH/BD/46711/ 2008.

**References**

[1] F. Grañena et al. (NEXT Collaboration), *NEXT Letter of Intent*, Laboratorio Subterráneo de Canfranc EXP-05 [hep-ex/0907.4054v1].

[2] D. Nygren, *Optimal detectors for WIMP and 0–ν ββ searches: Identical high-pressure xenon gas TPCs?*, Nucl. Instrum. Meth. A 581 (2007) 632.

[3] K. Yamamoto et al., *Development of Multi-Pixel Photon Counter (MPPC)*, IEEE Nuclear Science Symposium Conference Record (2007) 1511.

[4] J.A.M. Lopes et al., *A xenon gas proportional scintillation counter with a UV-sensitive, large-area avalanche photodiode*, IEEE Trans. Nucl. Sci. 48 (2001) 312.

[5] P.K. Lightfoot et al., *Optical readout tracking detector concept using secondary scintillation from liquid argon generated by a thick gas electron multiplier*, J. Inst. 4 (2009) P04002.

[6] Y. Giomataris, *Development and prospects of the new gaseous detector "Micromegas"*, Nucl. Instrum. Meth A 419 (1998) 239.

[7] A. Tomas et al., *Development of Micromegas for neutrinoless double beta decay searches*, J. Inst. 4 (2009) P11016.

[8] S. Andriamonje et al., *Development and performance of Microbulk Micromegas detectors*, J. Inst. 5 (2010) P02001.

[9] V. Aulchenko et al., *Further studies of GEM performance in dense noble gases*, Nucl. Instr. Meth. A 513 (2003) 256 and references therein.

[10] F.D. Amaro et al., *Operation of MHSP multipliers in high pressure pure noble-gas*, J. Inst. 1 (2006) P04003.

[11] A.S. Conceição, et al., *Operation of a single-GEM in noble gases at high pressures*, 2007 J. Inst. 2 P09010.

[12] R. Alon et al., *Operation of a Thick Gas Electron Multiplier (THGEM) in Ar, Xe and Ar-Xe*, J. Inst. 3 (2008) P01005.

[13] *Windowless Large Area APDs,* Application notes 1999, Advanced Photonix, Inc., 1240 Avenida Acaso, Camarillo, CA 93012, USA. http://www.advancedphotonix.com/

[14] J.M.F. dos Santos et al., *Development of portable gas proportional scintillation counters for X-ray spectrometry*, X-Ray Spectrom. **30** (2001) 373.

[15] S. Andriamonje et al, *Development and performance of Microbulk Micromegas detectors*, J. Inst. 5 (2010) P02001.

[16] L.M.P. Fernandes et al., *LAAPD low temperature performance in X-Ray and visible light detection,* IEEE Trans. Nucl. Sci. 51, 1575 (2004).

[17] R. Neilson, F. LePort, A. Pocara, K. Kumar, A. Odian et al. (EXO Collaboration), *Characterization of large area APDs for the EXO-200 detector*, Nucl. Instr. Meth. A **608** (2009) 68.

[18] C.M.B. Monteiro et al., *Detection of VUV photons with large area avalanche photodiodes*, Appl. Phys. B 81 (2005) 531.

[19] A. Delbart et al., *New developments of micromegas detector*, Nucl.Instrum.Meth. A 461 (2001) 84.

[20] C.M.B. Monteiro et al., *Secondary scintillation yield from gaseous micropattern electron multipliers in direct dark matter detection*, Phys. Lett. B 677 (2009) 133.

[21] A. Buzulutskov, *Physics of multi-GEM structures*, Nucl. Instr. Meth. A 494 (2002) 148, and references therein.

[22] A. Bondar et al., *Cryogenic avalanche detectors based on gas electron multipliers*, Nucl. Instr Meth. A 524 (2004) 130.

[23] T.H.V.T. Dias et al., *Full-energy absorption of X-ray energies near the Xe L-and K-photoionization thresholds in xenon gas detectors: Simulation and experimental results*, J. Appl. Phys. **82** (1997) 2742.

[24] B. Zhou, M. Szawlowski, *An explanation on the APD spectral quantum efficiency in the deep UV range*, Interoffice Memo, Advanced Photonix Inc., 1240 Avenida Acaso, Camarillo, CA 93012, EUA, 1999.

[25] J.A.M. Lopes et al., *A xenon gas proportional scintillation counter with a UV-sensitive, large-area avalanche photodiode*, IEEE Trans. Nucl. Sci., **48** (2001) 312-319.

[26] G.F. Knoll, Radiation Detection and Measurement, 3rd Edition, Wiley, New York, 2000.

[27] C.M.B. Monteiro et al, *Secondary scintillation yield in pure xenon*, J. Inst. 2 (2007) P05001.

[28] C.M.B. Monteiro et al, *Secondary scintillation yield in pure argon*, Phys. Lett. B 668 (2008) 167.

[29] L.C.C. Coelho, et al., *Xenon GPSC high-pressure operation with a large-area avalanche photodiode readout*, Nucl. Instr. Meth. A 575 (2007) 444.

[30] T. Dafni *et al*., *Energy resolution of alpha particles in a microbulk Micromegas detector at high pressure argon and xenon mixtures*, Nucl. Instrum. Meth. A 608, (2009) 259.

[31] S. Cebrián et al., *Micromegas readouts for double beta decay searches*, arXiv:1009.1827v1 [physics.ins-det] 9 Sep 2010.